\documentclass[english, 11pt]{article}
\PassOptionsToPackage{hyphens}{url}
\usepackage{subcaption}
\usepackage{fancyhdr}  
\usepackage{lastpage}  

\usepackage[margin=1in]{geometry}
\usepackage{lineno}

\usepackage[sorting=none, url=false, isbn=false, eprint=false]{biblatex}
\usepackage[utf8]{inputenc}
\usepackage{physics,amssymb}  

\usepackage{comment}
\usepackage{siunitx}
\usepackage{adjustbox}
\usepackage{csquotes}
\usepackage{xcolor}     
\usepackage{graphicx}
\usepackage{multirow}
\usepackage{hyperref}         
\usepackage{listings}         
\usepackage{algorithm}

\usepackage[title]{appendix}
\hypersetup{ 
    colorlinks,
    linkcolor={red!50!black},
    citecolor={blue!50!black},
    urlcolor={blue!80!black}}

\usepackage{hyperref}
\usepackage{amsmath}

\begin{document}
\let\WriteBookmarks\relax
\def\floatpagepagefraction{1}
\def\textpagefraction{.001}

\newcommand{\authcredits}[2]{%
    \noindent \textbf{#1:} #2
}

\newcommand{\keywords}[1]{%
  \vspace{1em}                     
  \noindent \textbf{Keywords:} #1 
}

\newcommand{\sectionstylee}[1]{\noindent{\normalfont\normalsize\sffamily\bfseries #1}}

\title{\fontsize{16pt}{16pt}\selectfont A geometric feature tracking approach for noninvasive patient specific estimation of leaflet strain from 3D images of heart valves}  

\author{\fontsize{12pt}{12pt}\selectfont Wensi Wu$^{1,2,}$\thanks{Corresponding author: wensiwu@seas.upenn.edu}, Matthew Daemer$^{3}$, Jeffrey A. Weiss$^{4,5}$, Alison M. Pouch$^{6}$, Matthew A. Jolley$^{3}$} 

\date{\fontsize{12pt}{12pt}\selectfont\today}

\maketitle

\begin{minipage}[t]{14cm}
    \small
  
    $^1$ Department of Mechanical Engineering and Applied Mechanics, University of Pennsylvania, Philadelphia, PA.  
    
    $^2$ Cardiovascular Institute, Children’s Hospital of Philadelphia, Philadelphia, PA.
    
    $^3$ Department of Anesthesiology and Critical Care Medicine, Children's Hospital of Philadelphia, Philadelphia, PA.
    
    $^4$ Department of Biomedical Engineering, University of Utah, Salt Lake City, UT.
    
    $^5$ Scientific Computing Institute, University of Utah, Salt Lake City, UT.
    
    $^6$ Department of Radiology and Bioengineering, University of Pennsylvania, Philadelphia, PA.

\end{minipage}


\begin{abstract}
\sectionstylee{Purpose} Valvular heart disease is prevalent and a major contributor to heart failure. Valve leaflet strain is a promising metric for evaluating the mechanics underlying the initiation and progression of valvular pathology. However, generalizable methods for noninvasively quantifying valvular strain from clinically acquired patient images remain limited. This study aims to develop a robust feature-tracking framework that enables accurate shape matching across variable valve morphologies and quantification of in vivo atrioventricular leaflet strain from three-dimensional echocardiographic (3DE) images in pediatric and adult patients. \\
\sectionstylee{Methods} We developed a geometric feature-tracking framework to quantify in vivo leaflet strain from 3DE images and to assess anatomical deformation across the cardiac cycle. The method integrates a cohort-derived geometric reference atlas to establish geometric correspondence and introduces a novel distance-weighted coherent point drift algorithm within a Gaussian mixture model framework for non-rigid registration. We evaluated performance against a finite element benchmark model and compared the approach with conventional point-based tracking methods. The framework was applied to pediatric and adult patient datasets (N = 31) to assess robustness across variable valve morphologies. \\
\sectionstylee{Results} The proposed method demonstrated greater accuracy in quantifying anatomical alignment and leaflet strain than conventional point-based approaches. Validation against the finite element benchmark confirmed improved strain estimation. The framework achieved reliable inter-phase tracking of valve deformation across diverse morphologies in pediatric and adult patients. Analysis identified a consistent distribution pattern of the $1^{\text{st}}$ principal strain associated with leaflet billow (prolapse). \\
\sectionstylee{Conclusion} This feature-tracking framework provides a generalizable method for noninvasive quantification of atrioventricular valve leaflet strain from clinical 3DE images. Characterization of biomechanical strain patterns may improve prognostic assessment and support longitudinal evaluation of valvular heart disease. Further investigation of the biomechanical signatures of heart valve disease has the potential to enhance prognostic assessment and longitudinal evaluation of valvular heart disease.
\end{abstract}


\maketitle

\section{Introduction}
Valvular heart disease (\textit{e.g.}, valve regurgitation or stenosis) is a prevalent cardiac condition that contributes substantially to the development of heart failure across all age groups~\cite{coffey_global_2021, benfari_excess_2019, messika-zeitoun_impact_2020, nishimura_mitral_2016, king_natural_2022}. Echocardiographic imaging is the clinical gold standard for evaluating morphological abnormalities of the heart valves (\textit{e.g.}, leaflet billowing, prolapse, and tenting)~\cite{mahmood_quantitative_2015}. 3D echocardiography (3DE) in particular enables visualization of dynamic 3D valve geometry, as well as quantification of functional metrics such as regurgitant volume and effective orifice area, which are critical indicators of disease severity~\cite{nam_dynamic_2022, nam_visualization_2022, nguyen_dynamic_2019}. However, global morphological features alone are not sufficient to characterize the biomechanical environment that governs disease initiation, progression, or the long-term durability of surgical repair. Strain analysis, in contrast, provides a detailed map of local tissue deformation that reflects tissue stiffness and applied boundary conditions. This information may serve as a proxy for identifying leaflet regions at increased risk of degeneration and potential tissue damage or rupture~\cite{lee_quantification_2015, zhang_meso-scale_2016, zhang_simulating_2021, sadeghinia_quantified_2024, schutte2025prediction}. Reliable non-invasive assessment of leaflet strain from 3DE images, transesophageal echocardiogram (TEE) and transthoracic echocardiogram (TTE), thus holds great potential for advancing prognostic evaluation of valvular disease and improving predictions of surgical repair durability~~\cite{ben_zekry_patient-specific_2016, el-tallawi_valve_2021}.

Assessment of leaflet strain is central to understanding the mechanical environment that drives tissue and cellular changes during valve disease initiation and progression~\cite{lee_quantification_2015, zhang_meso-scale_2016, zhang_simulating_2021, sadeghinia_quantified_2024}. Finite element analysis (FEA) has been widely used to investigate leaflet biomechanics and to examine the effects of annular dynamics~\cite{mathur_tricuspid_2019, larue_tricuspid_nodate}, papillary muscle displacement~\cite{spinner_effects_2012}, constitutive properties~\cite{wu_effects_2023, cai_effects_2019}, and annular geometry~\cite{salgo_effect_2002} on valve function~\cite{rego_noninvasive_2018, narang_pre-surgical_2021, gaidulis_patient-specific_2022, mathur_tricuspid_2019, larue_tricuspid_nodate, laurence_febio_2025, Bahadormanesh-Doppler-2026, pouch2012, Sadeghinia2023}. These studies have provided important insights into the fundamental mechanisms of valvular biomechanics. Despite these advances, the fidelity of FEA depends on how accurately the model reproduces the native valvular system. Most studies rely on animal, cadaveric, or in vitro data, and many incorporate tissue properties derived from non-human specimens. Even with image-derived FEA that enables patient-specific reconstruction from clinical imaging~\cite{kong_finite_2018, rego_noninvasive_2018, laurence_febio_2025, Bahadormanesh-Doppler-2026}, significant challenges remain. The spatial resolution of 3DE limits geometric accuracy. Complex leaflet morphology complicates fitting of finite element structures to image-derived models. These approaches often require model simplification and individualized adjustment of models, particularly for subvalvular chordal structures. Assumptions regarding surrogate chordal force representations are frequently necessary. The absence of a standardized and generalized framework for patient-specific valve modeling introduces user bias and variability, which may limit clinical reliability.

Feature tracking techniques, including image intensity–based feature tracking~\cite{heyde_elastic_2013, myronenko_intensity-based_2010, johnson_consistent_2002} and point–based feature tracking~\cite{beetz_modeling_2024, myronenko_point_2010}, present an alternative approach for estimating leaflet strain. Feature tracking approximates local tissue deformation by aligning geometric features between image frames. This characteristic eliminates the need for prior knowledge of material properties or chordal structure, making this approach well-suited for clinical translation. Notably, myocardial strain analysis has been widely adopted clinically to assess ventricular function. Multiple population studies suggested that myocardial strain analysis provides added value in improving diagnostic accuracy and prognostication in heart failure~\cite{namasivayam_utility_2023, erley_myocardial_2020, dong_incremental_2024, zhu_right_2025, shukla_prognostic_2018}. While feature tracking approaches hold considerable promise for valvular strain analysis, application of feature tracking-based approaches for valvular strain analysis remains in its infancy, with only a few notable studies to date~\cite{azencott_diffeomorphic_2010, el-tallawi_valve_2021, ben_zekry_patient-specific_2016}. 

Existing valvular feature-tracking studies have focused exclusively on mitral valves (MVs), with most algorithms being proprietary or limited by reliance on commercial software designed for MVs alone~\cite {azencott_diffeomorphic_2010, el-tallawi_valve_2021, ben_zekry_patient-specific_2016}. Due to challenges posed by large and rapid geometric changes in the leaflets during the transitions between open and closed time frames, prior work primarily tracked valves from mid-systole to end-systole, which does not capture the near strain free diastolic state of an open valve as a baseline. While 3D + time (4D) CT~\cite{williams2022truncal} and cardiac magnetic resonance imaging~\cite{iacovella2026rapid} are increasingly being utilized to evaluate valve function, there is currently no generalizable framework for analysis of valve leaflet strain across all 3D modalities. The need for multimodality-based assessment of valve leaflet strain across diverse valve morphologies and populations, including patients with complex congenital heart disease, is the key motivation for the present study.

To address these limitations, we sought to develop a robust open-source, feature tracking-based strain quantification framework for the assessment of diverse valve morphologies, including patients with congenital heart disease. Our method incorporates a novel distance-decay function into the Gaussian prior of the coherent point drift algorithm to robustly track geometries with substantial deformation~\cite{yuille_motion_1988, myronenko_point_2010}. The proposed method was verified using an FEA benchmark example, where it achieved more accurate shape and strain estimation than other published point-based feature tracking techniques. To further demonstrate robustness across a broad range of valve morphologies and types, we applied this verified framework to three patient cohorts, including adult and pediatric patients with congenital heart disease. Reliable feature tracking and strain quantification were achieved in both MVs and tricuspid valves (TVs) in patients across a broad age range, from infants to older adults with both normal and diseased morphologies.
\section{Materials and Methods}
\begin{figure*}[h!]
  \includegraphics[width=\linewidth]{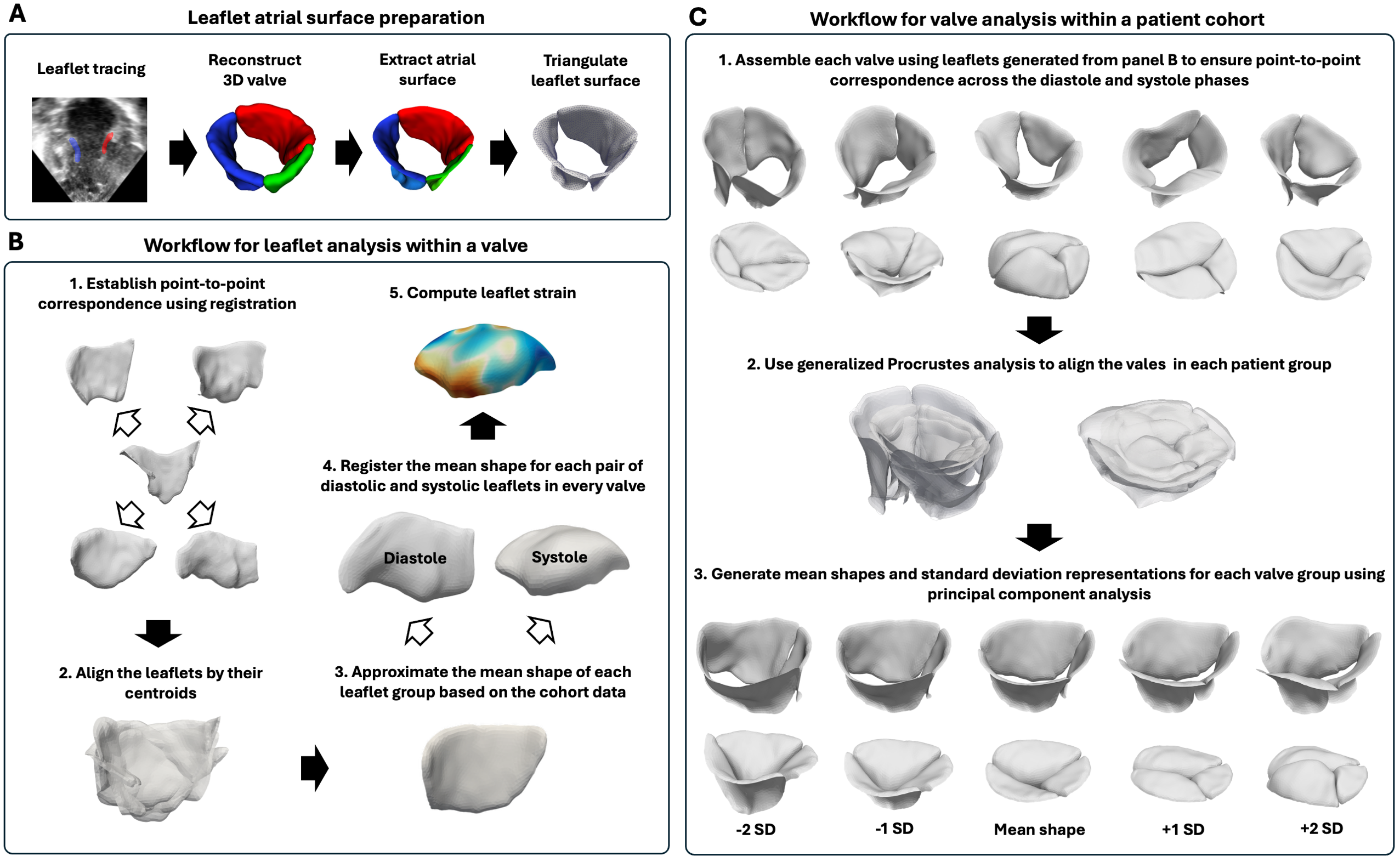}
  \caption{\textbf{Feature tracking-informed strain quantification framework.} (\textbf{A}) The atrial surfaces of the valves were extracted from segmentation models and triangulated with 1000 points uniformly distributed across each leaflet. (\textbf{B}) Within each patient cohort, one diastolic valve was randomly selected as the reference. Leaflets of the same anatomical category were registered to the corresponding reference leaflet to establish point-to-point correspondence. The registered leaflets were then aligned, and the mean shape (reference atlas) was estimated using generalized procrustes analysis. The mean shape was registered to both the diastolic and systolic frames of that leaflet and leaflet deformation was derived to compute strain on individual leaflets. (\textbf{C}) For patient cohort analysis, mean valve shapes at diastole and systole were first estimated using generalized procrustes analysis. Principal component analysis was applied to determine shape variation within each cohort. Finally, valves were registered between the open and closed phases to estimate systolic valve strain.}
  \label{fig:procedure}
\end{figure*}

Our proposed feature tracking-informed strain quantification framework is presented in Fig.~\ref{fig:procedure}. In all subjects, the areal and 1$^\text{st}$ principal strain at the mid-systolic frame were computed using the mid-diastolic frame as reference. The code of our strain calculation framework will be made available at \url{https://github.com/ww382/feature_tracking_informed_valve_strain}.

\subsection{Heart valve atrial surface processing}
Fig.~\ref{fig:procedure}A presents the process of preparing feature tracking-ready atrial surfaces of leaflets from 3DE images. 3DE images of pediatric TVs from patients with hypoplastic left heart syndrome (HLHS) (n = 11; 2 TEE and 9 TTE) and MVs (n = 10; 3 TEE and 7 TTE) were obtained from an existing database at the Children’s Hospital of Philadelphia. These images were acquired on a Philips Epiq system (Philips Medical, Boston, MA). In addition, transesophageal 3DE images from adult patients (n = 10) with normal MV morphology were obtained from an existing database at the University of Pennsylvania and were acquired on a Philips iE33 system (Philips Medical Systems, Andover, MA). Pre-existing segmentations for the authors' research groups were utilized. Since the image data originated from labs at different institutions, which had separate protocols for image export and manual segmentation, the pediatric data were imported into the SlicerHeart module~\cite{scanlan_comparison_2018, nguyen_dynamic_2019, lasso_slicerheart_2022} in 3D Slicer for segmentation, while the adult data were manually traced using ITK-SNAP~\cite{yushkevich_user-guided_2006}. The valve leaflets at the mid-diastolic and mid-systolic frames of the cardiac cycle were segmented by an expert observer to facilitate the 3DE-informed strain analysis. These were multi-label segmentations with a separate label for each leaflet. Image voxels corresponding to the leaflets were manually labeled and converted into 3D models, as described in~\cite{nam_visualization_2022, nam_dynamic_2022, nam_modeling_2023, aly_vivo_2021}. The collection and analysis of these data were approved by the Institutional Review Boards at the Children’s Hospital of Philadelphia and the University of Pennsylvania.

The atrial surface of each leaflet segmentation was extracted using the Leaflet Analysis tool in SlicerHeart. In particular, a closed curve was defined along the margin of the leaflet that separates the atrial and ventricular sides of the segmentation model. The model was then clipped along this curve to isolate the atrial surface. A custom Python script that utilized the PyACVD library~\cite{valette_generic_2008} was subsequently used to process the extracted surface. First, the extracted surface was cleaned by removing isolated and unreferenced points and then smoothed using Taubin smoothing~\cite{taubin1995curve}. Second, the diastolic and systolic atrial surfaces of each leaflet were discretized to linear triangular meshes by distributing 1,000 nodes uniformly using Voronoi clustering.

\subsection{Distance-weighted coherent point drift (DW-CPD) feature tracking}
The development of our DW-CPD feature tracking approach was inspired by the original CPD formulation for nonrigid point set feature tracking~\cite{myronenko_point_2010}. In the standard CPD framework, the fixed point set is represented as the centroids of a Gaussian mixture model (GMM) with isotropic covariances and uniform priors. The moving point set is transformed to the fixed point set through iterative expectation–maximization (EM) optimization to maximize the posterior correspondence probabilities (posterior probability function). The deformation field is regularized by the motion coherence theory~\cite{yuille_motion_1988}, which penalizes non-smooth variations in the velocity field. This ensures that neighboring points maintain coherent movement and preserve the topological structure of the point set.

While effective in many applications, the original CPD formulation can lead to unphysical global distortions in heart valve modeling, particularly when substantial changes in leaflet concavity between the open and closed configurations (\textit{e.g.,} in the case of a billowing leaflet). To improve the generalizability of CPD across diverse valve geometries, we introduce an adaptive, distance-weighted modification to the mixture model priors. In the new formulation, the prior probability of correspondence is increased for spatially proximate point pairs across the moving and target sets, while the likelihood of correspondence is weakened for more distant pairs. The motivation of this strategy reduces spurious global matches, mitigates large-scale geometric distortions, and thus yields more accurate leaflet tracking.

\subsubsection{Gaussian mixture models}
Consider two point clouds $\mathcal{X} = (\mathbf{x}_1,\ldots,\mathbf{x}_N)^T \in \mathbb{R}^{N\times 3}$ and 
$\mathcal{Y} = (\mathbf{y}_1,\ldots,\mathbf{y}_M)^T \in \mathbb{R}^{M\times 3}$, where $\mathcal{X}$ denotes the nodal coordinates of the mid-systolic mesh (fixed frame) with $N$ nodes and $\mathcal{Y}$ denotes the nodal coordinates of the mid-diastolic mesh (moving frame) with $M$ nodes within each valve. In standard CPD, points in $\mathcal{Y}$ were considered GMM centroids with uniform priors and equal, isotropic covariances. The resulting mixture density for a point $\mathbf{x}\in\mathcal{X}$ is
\begin{equation}
p(\mathbf{x}_n) = \frac{1}{M} \sum_{m=1}^{M} \frac{1}{(2\pi\sigma^2)^{D/2}} 
\exp\!\left(-\frac{\|\mathbf{x}_n-\mathbf{y}_m\|^2}{2\sigma^2}\right),
\end{equation}
with $D=3$ and $\sigma$ denotes the standard deviation of the Gaussian distribution. In this present work, all mid-systolic and mid-diastolic meshes have the same number of nodes. To account for uncertainty in point correspondence (e.g., noisy points or outliers), CPD introduces an additional uniform component with weight $w\in[0,1)$, yielding
\begin{equation}
\label{eq:gmm}
p(\mathbf{x}_n) = w\,\frac{1}{N} \;+\; (1-w)\,\frac{1}{M} \sum_{m=1}^{M} \frac{1}{(2\pi\sigma^2)^{D/2}} 
\exp\!\left(-\frac{\|\mathbf{x}_n-\mathbf{y}_m\|^2}{2\sigma^2}\right).
\end{equation}
The transformation $\mathbf{T}$ that maps the GMM centroids $\mathbf{y}_m$ to their deformed locations is estimated by minimizing the regularized negative log-likelihood function
\begin{equation}
\label{eq:log_likelihood}
E(\mathbf{T},\sigma^2) = -\sum_{n=1}^N \log p(\mathbf{x}_n) \;+\; \frac{\lambda}{2}\,\phi(\mathbf{T}),
\end{equation}
where $p(\mathbf{x}_n)$ is the mixture model likelihood in Eq.~\ref{eq:gmm}, $\lambda$ is a weighting constant, and $\phi(\mathbf{T})$ is a regularization function that enforces motion coherence and smoothness in the displacement field. Herein, $\mathbf{T} \in \mathbb{R}^{M \times D}$ is denoted as 
\begin{equation}\label{transformation_eqn}
\mathbf{T}(\mathcal{Y}_0) = \mathcal{Y}_0 + \mathbf{G}\mathbf{W}, 
\end{equation}
where $\mathcal{Y}_0$ is the initial nodal positions in the moving point set, $\mathbf{W} \in \mathbb{R}^{M \times D}$ is the deformation matrix of Gaussian kernel weights and $\mathbf{G} \in \mathbb{R}^{M \times M}$ is a symmetric Gram matrix with elements
\begin{equation}
g_{ij} = \exp \left(-\frac{1}{2}\left\|\frac{\mathbf{y}_{0i}-\mathbf{y}_{0j}}{\beta}\right\|^2\right),
\end{equation}
and $\beta$ is the width of the Gaussian kernel and $\mathbf{y}_{0i}$ is the $i^\text{th}$ of the initial point cloud. The regularization function $\phi(\mathbf{T})$ is expressed as 
\begin{equation}
\phi(\mathbf{T}) = \text{trace}(\mathbf{W}^T)\mathbf{G}\mathbf{W}. 
\end{equation}

\subsubsection{Expectation-maximization optimization}

CPD formulates point set feature tracking as a probabilistic alignment problem by modeling one point set as centroids of a Gaussian mixture model and estimating a transformation that maximizes the likelihood of the observed data. Since direct optimization of the log-likelihood is intractable, CPD employs the expectation-maximization (EM) algorithm to estimate soft correspondences and feature tracking parameters iteratively.

In the expectation (E) step, given estimates of the transformation $\mathbf{T}$ and the Gaussian variance $\sigma^2$, the posterior correspondence probability $P_{mn}$ between the transformed source point $\mathbf{T}(\mathbf{y}_m)$ and the target point $\mathbf{x}_n$ is computed as
\begin{equation}
P_{mn}= 
\frac{\exp \left(-\frac{\|\mathbf{x}_n-\mathbf{T}(\mathbf{y}_m)\|^2}{2\sigma^2}\right)}
{\sum_{k=1}^M\exp \left(-\frac{\|\mathbf{x}_n-\mathbf{T}(\mathbf{y}_k)\|^2}{2\sigma^2}\right)
+(2\pi\sigma^2)^{D/2}\frac{w}{1-w}\frac{M}{N}},
\end{equation}
where $P \in \mathbb{R}^{M \times N}$ denotes the soft posterior probability matrix and $w$ accounts for uniform noise and outliers.

In the proposed DW-CPD formulation, we introduce an additional distance-based weighting to encourage more localized correspondences. Specifically, the posterior probabilities are reweighted using an exponential distance decay function,
\begin{equation}
P_{mn} \propto P_{mn}\,\exp \left(-\gamma\|\mathbf{x}_n-\mathbf{T}(\mathbf{y}_m)\|\right),
\end{equation}
followed by normalization over $m$ for each $n$. The decay parameter $\gamma > 0$ controls the strength of the locality bias, with larger values favoring short-distance correspondences and smaller values allowing greater positional flexibility.

In the maximization (M) step, the nonrigid transformation $\mathbf{T}$ and the Gaussian variance $\sigma^2$ are updated in a sequential manner. Given the previous estimate of the variance $\sigma^2$ and the correspondence matrix $P$, the deformation matrix of Gaussian kernel weights $\mathbf{W} \in \mathbb{R}^{M \times D}$ is updated by solving the following linear system
\begin{equation}
\left(\mathbf{G} + \alpha\sigma^2\operatorname{diag}(P \mathbf{1})^{-1}\right)\mathbf{W}
=
\operatorname{diag}(P \mathbf{1})^{-1}P\mathcal{X} - \mathcal{Y},
\end{equation}
where $\mathbf{1}$ is a column vector of ones of length $N$, $\operatorname{diag}(P \mathbf{1})$ denotes the diagonal matrix whose diagonal entries are given by the vector $P \mathbf{1}$, and $\alpha$ is a regularization parameter enforcing motion coherence. This system admits a unique solution and yields an updated transformation $\mathbf{T}$ defined in Equation~\ref{transformation_eqn}.

With the updated transformation, the Gaussian variance $\sigma^2$ is subsequently updated analytically as
\begin{equation}
\sigma^2
=
\frac{1}{D\,\sum_{m,n} P_{mn}}
\sum_{m,n} P_{mn}
\left\|\mathbf{x}_n - \mathbf{T}(\mathbf{y}_m)\right\|^2,
\end{equation}
which estimates the residual alignment error between the transformed source points and the target points.

The EM procedure alternates between estimating soft correspondences in the E step and sequentially updating the transformation $\mathbf{T}$ and variance $\sigma^2$ in the M step to iteratively refine the nonrigid feature tracking. In the present work, we set $\alpha = 1$, $\beta = 5$, and $\gamma = 0.05$. Furthermore, we assume no outliers in the point clouds by setting $w=0$. 

\subsubsection{Feature tracking accuracy evaluation}
Feature tracking accuracy was evaluated by quantifying the geometric agreement between the registered valve surfaces and the reference geometries. The mean symmetric distance (MSD) and the 95$^{\text{th}}$ \%ile Hausdorff distance (HD$_{95}$) were used to evaluate shape similarity. Let $\mathcal{S}_1$ and $\mathcal{S}_2$ denote the registered surface and the reference surface, respectively, each represented as a discrete set of surface points. For a point $\mathbf{p} \in \mathcal{S}_1$, the point-to-surface distance to $\mathcal{S}_2$ is defined as
\begin{equation}
d(\mathbf{p}, \mathcal{S}_2) = \min_{\mathbf{q} \in \mathcal{S}_2} \|\mathbf{p} - \mathbf{q}\|,
\end{equation}
where $\mathbf{q}$ denotes a point on the surface $\mathcal{S}_2$.  
The MSD is then computed as
\begin{equation}
\mathrm{MSD}(\mathcal{S}_1, \mathcal{S}_2) =
\frac{1}{2}
\left(
\sum_{\mathbf{p} \in \mathcal{S}_1} d(\mathbf{p}, \mathcal{S}_2)
+
\sum_{\mathbf{q} \in \mathcal{S}_2} d(\mathbf{q}, \mathcal{S}_1)
\right).
\end{equation}
HD$_{95}$ is computed as,
\begin{equation}
\mathrm{HD}_{95}(\mathcal{S}_1, \mathcal{S}_2) =
\max\left\{
Q_{0.95}\!\left(\{d(\mathbf{p}, \mathcal{S}_2)\}_{\mathbf{p}\in\mathcal{S}_1}\right),
Q_{0.95}\!\left(\{d(\mathbf{q}, \mathcal{S}_1)\}_{\mathbf{q}\in\mathcal{S}_2}\right)
\right\},
\end{equation}
where $Q_{0.95}(\cdot)$ denotes the 95$^{\text{th}}$ \%ile of the distribution.

\subsection{3DE-informed strain analysis in vivo}\label{sec:strain}
Here, we present our workflow for strain analysis on individual leaflets (Fig.~\ref{fig:procedure}B) as well as cohort-based analysis (Fig.~\ref{fig:procedure}C) based on nodal correspondences obtained from DW-CPD.

\subsubsection{Individual lealfet analysis}
Leaflet geometry in atrioventricular valves exhibits substantial inter-subject variability, particularly in congenital heart disease~\cite{nam_modeling_2023}. To enable consistent feature tracking across subjects and cardiac phases, we constructed a representative mean leaflet shape for each anatomical label. Individual leaflets of the same label (e.g., anterior, posterior, septal) were first aligned using the VTK Python module by translating and rotating each leaflet such that its centroid shared a common reference point and the atrial surfaces faced the same orientation. The aligned leaflets were manually inspected to ensure that annular and free edges were oriented consistently across samples. A representative mid-diastolic leaflet from a randomly selected subject was then chosen as a reference, and all other mid-diastolic leaflets of the same anatomical label were registered to this reference using DW-CPD to establish intra-subject point correspondences. From these registered shapes, a mean leaflet geometry was computed using generalized Procrustes analysis. The resulting mean shape serves as a reference atlas that facilitates interpolation and intermediate transitioning of leaflet geometry between the open and closed configurations. This intermediate representation is critical for robust feature tracking between mid-diastolic and mid-systolic frames, particularly in pathological valves (e.g., billowing leaflets), where large nonrigid deformations and pronounced changes in leaflet concavity occur. The mean shape is computed only once; thereafter, it is used to facilitate analysis of all patients' valves.

The mid-diastolic and mid-systolic point sets were independently registered to the mean shape to establish consistent point correspondences across cardiac phases. The deformation gradient and strain were subsequently computed from the corresponding nodal coordinates between the two phases. Both areal strain and the 1$^\text{st}$ principal Green-Lagrange strain were considered in the present work. Areal strain provides an intuitive quantification of local in-plane surface expansion and compression during leaflet motion. In contrast, the 1$^\text{st}$ principal Green-Lagrange strain enables identification of regions experiencing elevated in-plane tensile deformation that may be susceptible to tissue damage or failure.

\textit{Areal strain.} For a triangular element on the atrial surface with vertices $\{\mathbf{x}_a\}_{a=1}^3$, the surface is parameterized on a reference triangle using local coordinates $(\xi,\eta)$ and linear shape functions
\begin{equation}
N_1(\xi,\eta) = \xi, \qquad
N_2(\xi,\eta) = \eta, \qquad
N_3(\xi,\eta) = 1 - \xi - \eta,
\end{equation}
which satisfy $\sum_{a=1}^3 N_a = 1$. The Cartesian coordinates of the surface are interpolated as
\begin{equation}
\mathbf{x}(\xi,\eta) = \sum_{a=1}^3 N_a(\xi,\eta)\,\mathbf{x}_a.
\end{equation}
Here, $\mathbf{x}_a \in \mathbb{R}^3$ denotes the $a^\text{th}$ vertex of a triangular element, written as a column vector $\mathbf{x}_a = [x_a,\, y_a,\, z_a]^\top$.

The surface Jacobian is defined as
\begin{equation}
\mathbf{J} =
\begin{bmatrix}
\frac{\partial \mathbf{x}}{\partial \xi} \\
\frac{\partial \mathbf{x}}{\partial \eta}
\end{bmatrix}
=
\begin{bmatrix}
\frac{\partial N_1}{\partial \xi} & \frac{\partial N_2}{\partial \xi} & \frac{\partial N_3}{\partial \xi} \\
\frac{\partial N_1}{\partial \eta} & \frac{\partial N_2}{\partial \eta} & \frac{\partial N_3}{\partial \eta}
\end{bmatrix}
\begin{bmatrix}
\mathbf{x}_1^\top \\
\mathbf{x}_2^\top \\
\mathbf{x}_3^\top
\end{bmatrix},
\end{equation}
where the derivatives of the shape functions are constant over each element. The surface Jacobian $\mathbf{J}$ yield two surface tangent vectors,
\begin{equation}
\mathbf{t}_1 = \frac{\partial \mathbf{x}}{\partial \xi},
\qquad
\mathbf{t}_2 = \frac{\partial \mathbf{x}}{\partial \eta}.
\end{equation}
The element area is then evaluated as
\begin{equation}
A = \frac{1}{2}\|\mathbf{t}_1 \times \mathbf{t}_2\|.
\end{equation}

For each triangulated surface element, the surface area change is quantified using the areal strain,
\begin{equation}
E_{\text{areal}} = \frac{A_{\text{def}} - A_{\text{ref}}}{A_{\text{ref}}},
\end{equation}
where $A_{\text{ref}}$ and $A_{\text{def}}$ denote the element areas in the reference (mid-diastolic) and deformed (mid-systolic) configurations, respectively.

\textit{Green-Lagrange strain.} To compute in-plane strain, a local deformation gradient is estimated by mapping the reference element basis to the corresponding deformed element basis. For each triangular element, two surface tangent vectors, $\mathbf{t}_1$ and $\mathbf{t}_2$, are first obtained from the surface Jacobian. The element normal is then computed as
\begin{equation}
\mathbf{n} = \frac{\mathbf{t}_1 \times \mathbf{t}_2}{\|\mathbf{t}_1 \times \mathbf{t}_2\|}.
\end{equation}
Using the tangent and normal vectors, the $3\times3$ surface basis matrices for the reference and deformed configurations are constructed as
\begin{equation}
\mathbf{B}_{\text{ref}} =
[\mathbf{t}_{1,\text{ref}} \ \mathbf{t}_{2,\text{ref}} \ \mathbf{n}_{\text{ref}}],
\qquad
\mathbf{B}_{\text{def}} =
[\mathbf{t}_{1,\text{def}} \ \mathbf{t}_{2,\text{def}} \ \mathbf{n}_{\text{def}}],
\end{equation}
where $\mathbf{t}_{1}$ and $\mathbf{t}_{2}$ are the surface tangents obtained from the Jacobian, and $\mathbf{n}$ is the corresponding unit normal.

The deformation gradient mapping the reference configuration to the deformed configuration is computed as
\begin{equation}
\mathbf{F} = \mathbf{B}_{\text{def}}\,\mathbf{B}_{\text{ref}}^{-1}.
\end{equation}
The right Cauchy--Green tensor and Green-Lagrange strain tensor are then defined as
\begin{equation}
\mathbf{C} = \mathbf{F}^\top \mathbf{F},
\qquad
\mathbf{E}_{\text{Green--Lagrange}} = \frac{1}{2}(\mathbf{C}-\mathbf{I}),
\end{equation}
where $\mathbf{I}$ is the $3\times3$ identity tensor.

\subsubsection{Cohort-based whole valve analysis}
With point-to-point correspondence established across all valves, we performed a cohort-based shape analysis to characterize morphological variability within each patient cohort and to provide geometric context for the interpretation of leaflet strain. The objectives of this analysis were to construct statistical shape models of atrioventricular valve geometry at mid-diastole and mid-systole, quantify the dominant modes of geometric variation within each cohort, and relate these variations to observed strain patterns.

Valve geometries at each cardiac phase (mid-diastole and mid-systole) were analyzed separately. Within each phase, all valve shapes were rigidly aligned using generalized Procrustes analysis such that they share the same origin, orientation, and scale. A mean valve geometry was then computed for each cohort and cardiac phase. Owing to substantial inter-subject variability in valve anatomy and orientation, particularly within the HLHS tricuspid valve cohort, automatic alignment occasionally resulted in inconsistent leaflet orientations. In such cases, the misaligned valves were manually rotated to ensure consistent anatomical orientation prior to mean shape computation.

Principal component analysis (PCA) was subsequently performed on the aligned valve geometries at each cardiac phase to quantify shape variability within each patient cohort. Each valve shape was represented as a high-dimensional vector formed by concatenating the three-dimensional coordinates of all mesh nodes. The resulting principal components define orthogonal modes of geometric variation about the mean shape. Shape variation along each mode was achieved by reconstructing shapes at $\pm1$ and $\pm2$ standard deviations (SD) from the mean according to
\begin{equation}
\mathbf{x}_{\pm k} = \bar{\mathbf{x}} \pm k\sqrt{\lambda_i}\,\mathbf{v}_i.
\end{equation}
Here, $k$ denotes the number of standard deviations along the $i^\text{th}$ principal component used to reconstruct shape variation, $\bar{\mathbf{x}}$ denotes the mean shape, and $\lambda_i$ and $\mathbf{v}_i$ are the eigenvalue and eigenvector associated with the $i^\text{th}$ principal component.

To evaluate leaflet deformation and strain between cardiac phases, diastolic and systolic valve geometries for each subject were then registered using DW-CPD, and areal and Green-Lagrange strains were computed on the resulting corresponding meshes. As a result, strain patterns can be further analyzed within the context of inter-subject morphological variability.
\section{Results}
We performed comprehensive leaflet strain analyses on three patient cohorts from the Children's Hospital of Philadelphia and the University of Pennsylvania. Detailed verification results on shape agreement using our proposed method, together with a comparative analysis against other point-based feature tracking techniques, are provided in Section~\ref{sec:verification}. Section~\ref{sec:HLHS_TV_strain} presents strain analysis of TVs in patients with hypoplastic left heart syndrome (HLHS). Section~\ref{sec:pediatric_MV_strain} provides strain analysis of pediatric MVs, including both normal and a range of pathologic conditions, such as rheumatic mitral regurgitant valve and Marfan syndrome. Section~\ref{sec:adult_MV_strain} covers the strain analysis of normal adult MVs. In this context, ``normal" indicates valves that coapt properly without regurgitation or tissue calcification. However, these adult patients had other cardiac conditions, such as aortic stenosis or coronary artery disease, at the time of imaging. All feature tracking-based strain analyses were performed on a MacBook Pro with an Apple M2 Max chip (Apple Computer, Cupertino, CA). The average computational time for feature tracking and strain estimation combined was 14.2 seconds per HLHS tricuspid valve, using meshes with 1000 nodes per leaflet.

\subsection{Finite element analysis verification}~\label{sec:verification}
An image-derived mitral valve (MV) model from Wu et al.\cite{wu_computational_2022} was used as the benchmark to verify the proposed algorithm. The ground truth valve deformation, areal strain, and 1$^\text{st}$ principal Green Lagrange strain were generated in FEBio using the same constitutive model parameters and pressure. The undeformed (mid-diastolic) mesh and the deformed (mid-systolic) mesh were used as inputs to all feature tracking methods to evaluate their performance in approximating the deformed geometry and leaflet surface strain relative to FE–derived approximations. The strain formulation is described in Section~\ref{sec:strain}. We compared the feature tracking and strain estimation performance of DW-CPD against established point-based feature tracking approaches: Bayesian-CPD (BCPD)~\cite{hirose_bayesian_2021} and classic CPD~\cite{myronenko_point_2010}. Probreg~\cite{probreg} was used to facilitate the BCPD feature tracking. 

\begin{figure*}[h!]
\centering
\includegraphics[width=1\textwidth]{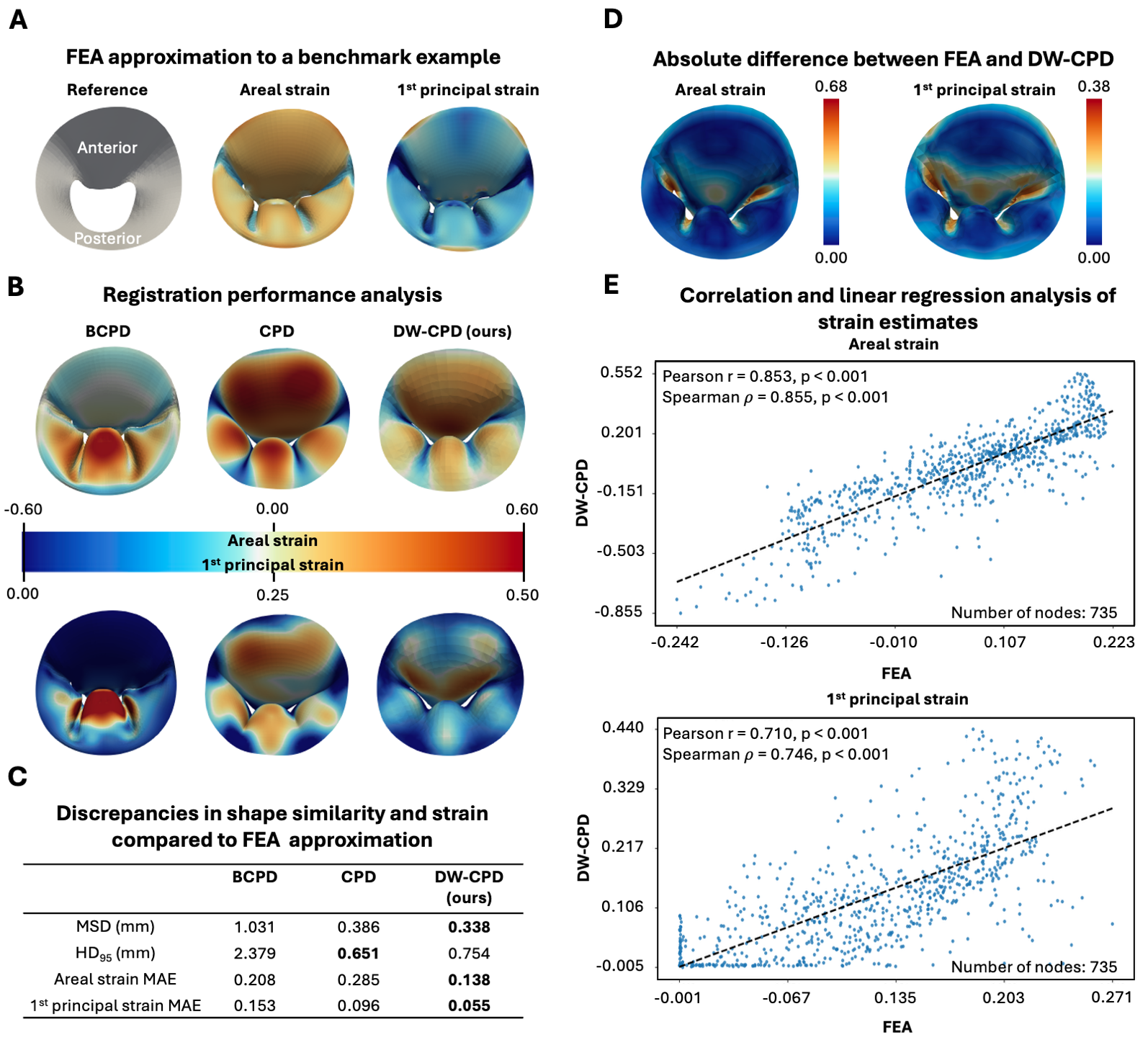}
\caption{\textbf{Feature tracking verification results.} (\textbf{A}) Ground truth areal and 1$^\text{st}$ principal strain of an image-derived MV finite element model were used to compare the performance of the proposed and established feature-tracking methods. (\textbf{B}) Qualitative comparisons of the deformed geometry and strain distributions are provided. Both CPD and DW-CPD successfully captured the large valve deformation and the strain patterns. (\textbf{C}) DW-CPD yielded more accurate estimates of mean symmetric distance, area strain MAE, and 1$^{\text{st}}$ principal strain MAE compared to existing CPD approaches. (\textbf{D}) The discrepancy in strain estimation using DW-CPD is most noticeable in the commissural folds. (\textbf{E}) Scatter plots show strong correlations between areal and1$^{\text{st}}$ principal strain generated from Dw-CPD and FEA.} \label{fig:verification}
\end{figure*}

The reference areal and 1$^\text{st}$ principal strain are shown in Fig.~\ref{fig:verification}A. Fig.~\ref{fig:verification}B provides a qualitative comparison between the estimated and ground truth strain maps. As illustrated, although all feature tracking methods capture the deformed systolic shape to a large extent, noticeable differences are observed in the estimated strain distributions across the three methods. BCPD substantially underpredicts strain on the anterior leaflet of the mitral valve compared with CPD and DW-CPD. Both CPD and DW-CPD capture strain across the entire valve. However, CPD tends to overpredict strain in regions experiencing high tensile bending. Shape agreement was assessed using the MSD and HD$_{95}$, while strain accuracy was evaluated using the mean absolute error (MAE) of areal strain and first principal strain. Differences in the estimated systolic valve surface geometry and leaflet strain distributions are shown in Fig.~\ref{fig:verification}C. Both CPD and DW-CPD achieved substantially lower MSD and HD$_{95}$ than BCPD, with more than a three-fold improvement in shape similarity, and DW-CPD yielded the lowest geometric error. In addition, DW-CPD produced an areal strain MAE of 0.138 and a 1$^\text{st}$ principal strain MAE of 0.055, the lowest among all methods evaluated.

Visualizations of differences in areal strain and 1$^\text{st}$ principal strains between DW-CPD and FEA are shown in Fig.~\ref{fig:verification}D. The DW-CPD appears to overpredict compressive and tensile strain in high curvature regions, especially near commissure folds. Fig.~\ref{fig:verification}E presents scatter plots of areal strain and $1^{\text{st}}$ principal strain estimated by DW-CPD and FEA. The results demonstrate strong agreement between the two methods. For areal strain, the correlation was high, with Pearson’s $r = 0.853$ ($p < 0.001$) and Spearman’s $\rho = 0.855$ ($p < 0.001$). For $1^{\text{st}}$ principal strain, Pearson’s $r$ was $0.710$ ($p < 0.001$) and Spearman’s $\rho$ was $0.746$ ($p < 0.001$).

\subsection{Segmental leaflet atrial surface evaluation}
Table~\ref{tab:leaflet_area} summarizes the surface area of the leaflets during mid-diastole and mid-systole for each segmental atrial surface across the three patient cohorts, along with the corresponding percent differences in surface area between the two phases. The differences between diastolic and systolic surface areas varied, showing both increases and decreases across leaflets and patients. This resulted in a wide distribution of percent differences. In the HLHS TVs, leaflet surface area changes were mixed, with several showing positive percent differences while others had values close to zero or negative. A similar variability was noted in the pediatric MVs. In the adult MVs, negative percent differences were more frequently observed in both the anterior and posterior leaflets, although some positive changes in surface area were still present.

\begin{table*}[h!]
    \centering
    \caption{\textbf{Leaflet surface area across cohorts.} (A) HLHS TVs present a combination of positive and negative changes in surface area between the anterior, posterior, and septal leaflets when comparing mid-diastole and systole frames. (B) Pediatric MVs similarly demonstrate a range of changes in surface area. (C) The percentage differences in leaflet surface area for adult MVs were mostly negative across both leaflets.}
    \label{tab:leaflet_area}
    \includegraphics[width=1\textwidth]{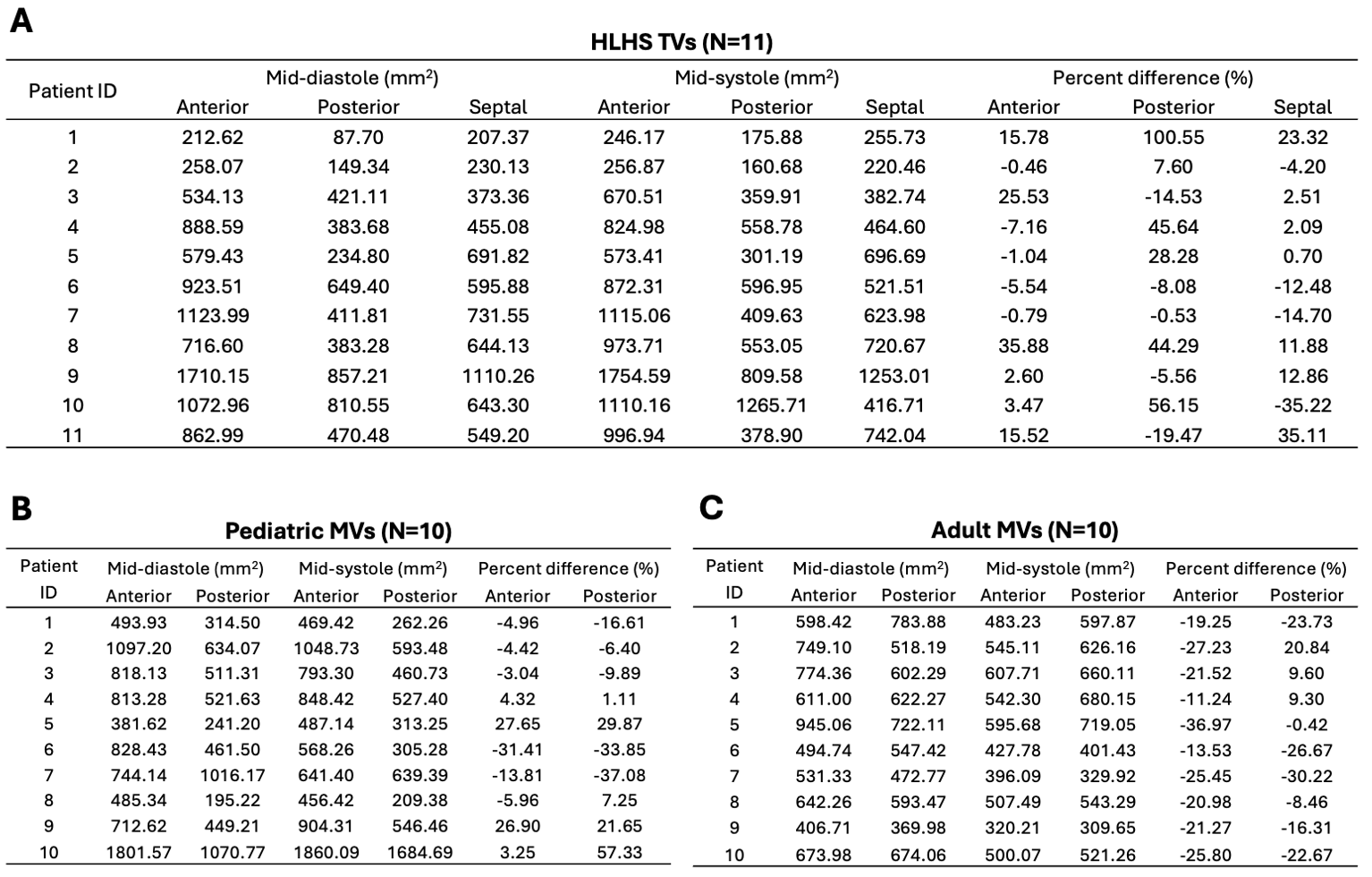}
\end{table*}

\subsection{Quantitative assessment of feature-tracking accuracy}
\begin{figure*}[h!]
\centering
\includegraphics[width=0.5\textwidth]{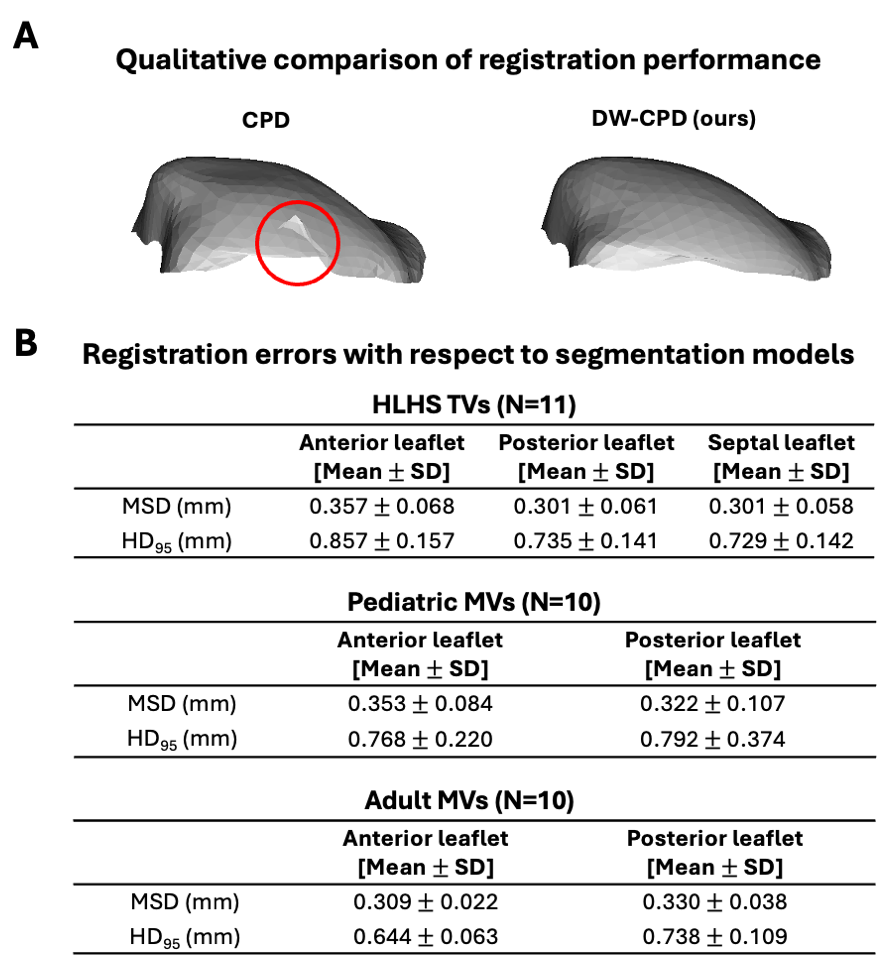}
\caption{\textbf{Feature tracking performance.} (\textbf{A}) This example illustrates that CPD resulted in self-penetration at the leaflet edge, while the proposed DW-CPD eliminated this defect with improved feature tracking fidelity. (\textbf{B}) DW-CPD was applied to 3 groups of atrioventricular valves and verified against ground truth segmentation. Strong shape agreement was observed, with MSD values below 0.4~mm and HD$_{95}$ values below 1~mm across all leaflets.}\label{fig:reg_performance}
\end{figure*}

We applied the proposed DW-CPD feature-tracking method to estimate leaflet deformation in three patient cohorts: (1) TV in patients with HLHS, (2) pediatric MVs, and (3) adult MVs. Fig.~\ref{fig:reg_performance}A shows an example where the CPD method produced a feature tracking artifact at the leaflet edge, causing it to fold inward and protrude, resulting in self-penetration. In contrast, DW-CPD accurately tracked the leaflet deformation, eliminating this defect. The mean and SD of the MSD and HD$_{95}$ for each leaflet within each cohort are summarized in Fig.~\ref{fig:reg_performance}B. DW-CPD was used to inform the strain calculation among the patient cohorts. Overall, the registered leaflets achieved strong agreement compared with the ground truth segmentation models. Among the patient groups, the anterior leaflet of HLHS TVs showed the largest errors, with an average MSD of 0.357mm $\pm$ 0.068 and HD$_{95}$ of 0.857~mm $\pm$ 0.157.  While our primary goal was to demonstrate that our methods could capture leaflet geometry across diverse valve types and morphologies, and we were not powered to determine granular associations of strain to morphology, we explored descriptive statistics of leaflet geometry and leaflet strain across all three patient populations as described in the following section.  This serves as a demonstration of how our methods could be applied to investigate associations of strain to valve disease and morphology in future studies, and to provide initial insights in pediatric patients where understanding of the relationships between valve morphology, strain, and pathology is nascent.

\newpage
\subsection{Strain assessment of pathological tricuspid valves (N=11)}~\label{sec:HLHS_TV_strain}

\begin{figure*}[h!]
\centering
\includegraphics[width=\textwidth]{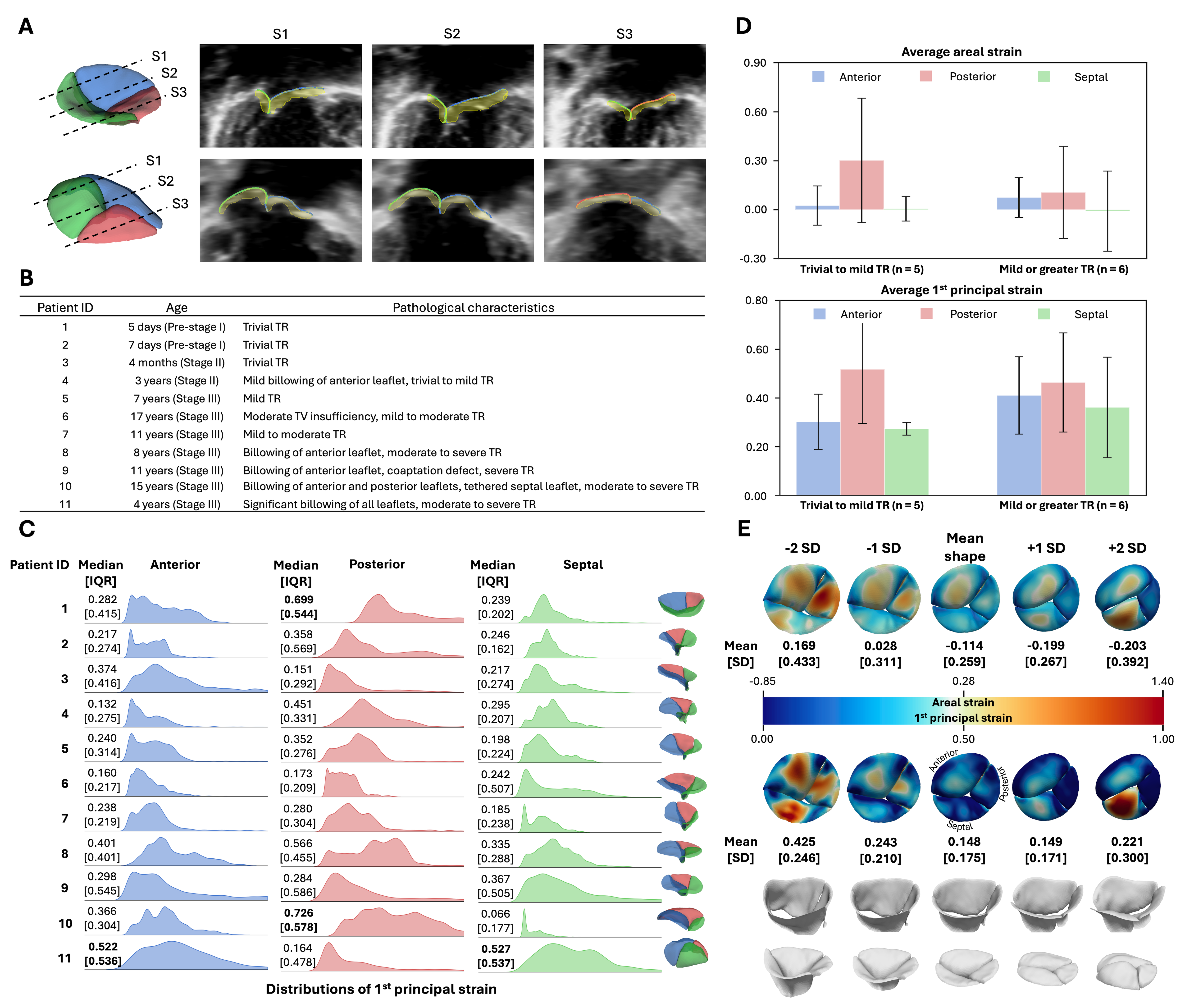}
\caption{\textbf{Pediatric tricuspid valve analysis (N=11).} (\textbf{A}) The estimated systolic atrial surfaces were superimposed onto 3D TEE images along with the ground truth segmentation models. Excellent feature tracking accuracy was obtained, as demonstrated in the visualization of the three cut planes shown. (\textbf{B}) The study cohorts included patients spanning a wide age range, from 5 days to 17 years. For each patient, the surgical stage at the time of image acquisition, as well as the pathological characteristics of the TVs, were reported. (\textbf{C}) The distributions of 1$^{\text{st}}$ principal strain, as well as their median and IQRs, were compared across the three leaflets in all 11 patients. Billowing leaflets were typically characterized by a high median and strain IQR, with a relatively uniform strain distribution profile. (\textbf{D}) Average areal strain and 1$^{\text{st}}$ principal strain were compared between the trivial to mild and mild or greater TR cohorts. The trivial to mild cohort exhibited lower areal and 1$^{\text{st}}$ principal strain in the anterior leaflet but higher values in the posterior leaflet relative to the mild or greater cohort. Both cohorts showed near zero areal strain in the septal leaflet, while the mild or greater cohort demonstrated higher 1$^\text{st}$ principal strain in the septal leaflet. (\textbf{E}) Population-based strain analysis was performed. 1$^\text{st}$ principal strain and IQRs increase as the valve morphology deviates from the mean shape.} \label{fig:pediatric_TV_analysis}
\end{figure*}
\subsubsection{Patient population} 
 This study cohort comprises 11 pediatric patients diagnosed with HLHS: 2 patients with HLHS prior to stage I palliation, 2 with stage II HLHS, and 7 with stage III HLHS following Fontan palliation. The patient cohort was randomly selected from our database. Patients' age at the time of image acquisition ranges from 4 months to 17 years. Of the 11 patients, 5 had trivial to mild tricuspid regurgitation (TR), and 6 developed mild or greater TR. Examples of the tracked leaflet atrial surface, along with segmentation ground truth, overlaid on 3DE images, are shown in Fig.~\ref{fig:pediatric_TV_analysis}A. Descriptions of the TV pathological characteristics (drawn from clinical reports) and the age associated with each patient are summarized in Fig.~\ref{fig:pediatric_TV_analysis}B. 

\subsubsection{Strain characteristics on individual leaflet} 
The areal and 1$^{\text{st}}$ principal strains at the mid-systole frame were calculated for each TV. Fig.~\ref{fig:pediatric_TV_analysis}C shows ridgeline plots of the 1$^\text{st}$ principal strain distributions for each leaflet in each patient, along with the corresponding median and interquartile range (IQR). Qualitatively, billowing leaflets showed a broader and flatter strain distribution, while valves with trivial to mild TR exhibited more sharply defined peaks and a narrower spread. This pattern indicates that strain in valves with trivial to mild TR is more localized, while strain in billowing leaflets is more uniformly distributed. Additionally, we observed that valves with a higher regurgitant grade tend to be accompanied by higher median and IQR of strain. The leaflets with both mean and IQR values greater exceed 0.5 are highlighted in bold. Of the 4 leaflets identified, 3 were confirmed as billowing.

We  compared the average areal and 1$^{\text{st}}$ principal strain in the trivial to mild TR and mild or greater TR cohorts. The mean and standard deviation (SD) of the average areal strain and 1$^{\text{st}}$ principal strain on each leaflet are shown in Fig.~\ref{fig:pediatric_TV_analysis}D. The average areal strain quantifies the relative change in leaflet surface area between diastolic and systolic frames, whereas the average 1$^{\text{st}}$ principal strain represents the mean maximum tensile strain across the leaflet. With respect to areal strain, both groups exhibited negligible changes in septal leaflet area while demonstrating increased anterior and posterior surface area. Both the trivial to mild TR and mild to greater TR cohorts showed similar anterior areal strain, although the trivial to mild TR demonstrated slightly lower values (0.025 $\pm$ 0.120 vs. 0.075 $\pm$ 0.124). However, posterior areal strain trended higher in the trivial to mild TR cohort compared to the mild or greater cohort (0.302 $\pm$ 0.381 vs. 0.106 $\pm$ 0.282). This suggests that  patients in the trivial and mild cohort may have more extensible posterior leaflets. Consistent with the areal strain findings, the trivial to mild TR cohort exhibited lower average 1$^{\text{st}}$ principal strain in the anterior leaflet compared to the mild or greater cohort (0.303 $\pm$ 0.113 vs. 0.410 $\pm$ 0.158), but higher average 1$^{\text{st}}$ principal strain in the posterior leaflet (0.517 $\pm$ 0.221 vs. 0.463 $\pm$ 0.203). For the septal leaflet, the average 1$^{\text{st}}$ principal strain was 0.273 $\pm$ 0.026 in the trivial to mild cohort and 0.361 $\pm$ 0.206 in the mild or greater cohort.

\subsubsection{Population study of strain patterns across the patient cohort}
Principal component analyses (PCA) of the TVs at mid-diastole and mid-systole were performed independently to determine the mean valve shape and its standard deviations (SD) across the 11 patients. Leaflet strains were computed by tracking point-to-point correspondences between the open and closed valve geometries using our distance-weighted coherent point drift (DW-CPD) feature tracking approach. The resulting spectrum of valve morphologies and associated strain information is shown in Fig.~\ref{fig:pediatric_TV_analysis}E. Within this spectrum, the valve appeared most tethered at –2 SD from the mean shape and most billowing at +2 SD. Both ends of the morphological spectrum demonstrated markedly elevated strains relative to the mean shape. In the most tethered valve, high strain was distributed across all three leaflets, whereas in the most billowing valve, high strain was concentrated on the septal leaflet. 

The mean shape exhibited relatively low strains (–0.114 $\pm$ 0.259 average areal strain and 0.148 $\pm$ 0.175 average 1$^{\text{st}}$ principal strain). In contrast, the most tethered valve showed an average areal strain of 0.169 $\pm$ 0.433 and a 1$^{\text{st}}$ principal strain of 0.425 $\pm$ 0.246, while the most billowing valve demonstrated an average areal strain of –0.203 $\pm$ 0.392 and a 1$^{\text{st}}$ principal strain of 0.221 $\pm$ 0.300. For both areal and 1$^{\text{st}}$ principal strain, the SD of the strain distribution increased as valve geometries deviated further from the mean shape. These findings suggest that both elevated 1$^{\text{st}}$ principal strain and greater strain variability, represented by SD, can be associated with pathological TV morphology.

\subsection{Strain assessment of normal and diseased pediatric mitral valves (N=10)}~\label{sec:pediatric_MV_strain}
\begin{figure*}[h!]
\centering
\includegraphics[width=\textwidth]{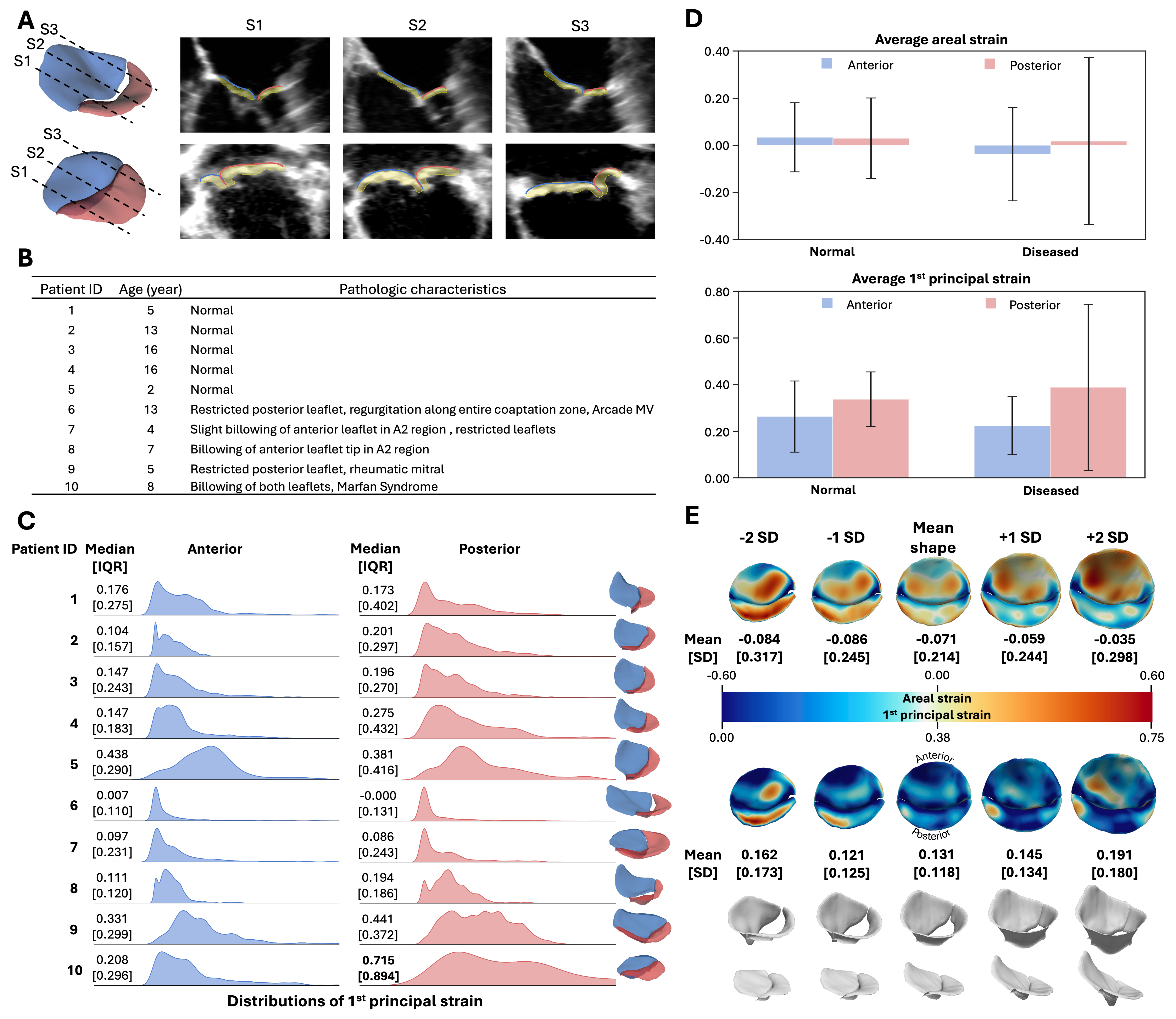}
\caption{\textbf{Pediatric mitral valve analysis (N=10).} (\textbf{A}) Visualization of the ground truth segmentation models and estimated atrial surfaces were superimposed onto 3D TEE images on the three cut planes shown. (\textbf{B}) The study cohorts included patients from 2 to 16 years old at the time of image acquisition. The MV pathological characteristics for each patient were provided. (\textbf{C}) The median, IQR, and distribution of 1$^{\text{st}}$ principal strain were compared across the anterior and posterior leaflets in all 10 patients. Patient 10 was the only case with leaflet strain exhibiting a median and IQR greater than 0.5, which also corresponded to billowing morphology. (\textbf{D}) Average areal strain and 1$^{\text{st}}$ principal strain were compared between the normal and diseased cohorts. Both cohorts showed negligible changes in leaflet surface area and demonstrated higher average 1$^{\text{st}}$ principal strain in the posterior leaflet than in the anterior leaflet. (\textbf{E}) Population-based strain analysis showed that higher average 1$^{\text{st}}$ principal strains at both ends of the shape spectrum compared with the mean shape. In addition, strain SD increased as valve morphology deviated from the mean.} \label{fig:pediatric_MV_analysis}
\end{figure*}

\subsubsection{Patient population}
This study cohort included 10 pediatric patients, aged 2 to 16 years at the time of image acquisition. Of these, 5 had normal coaptation, 3 demonstrated billowing of one or more leaflets, and 2 had restricted posterior leaflets. Further, the cohort included 1 patient with an MV arcade, 1 with rheumatic MV, and 1 with Marfan syndrome.  Excellent feature tracking accuracy was obtained, as demonstrated in the visualization of the three cut planes shown. Fig.~\ref{fig:pediatric_MV_analysis}A shows the superimposition of ground truth segmentation and the estimated atrial leaflet surface onto 3D TEE images from two representative patients. The ages and clinical diagnoses of all patients are summarized in Fig.~\ref{fig:pediatric_MV_analysis}B.

\subsubsection{Strain characteristics on individual leaflet} 
Ridgeline plots comparing the distributions, median and IQRs of 1$^{\text{st}}$ principal strain for the anterior and posterior leaflets are shown in Fig.~\ref{fig:pediatric_MV_analysis}C. The diagnoses among MV patients in this cohort were highly heterogeneous, resulting in considerable variability in strain distribution patterns. We highlighted the posterior leaflet of Patient 10, which showed a median strain and IQR greater than 0.5. Notably, the clinical note described this leaflet as billowing. This finding further suggests that the median and IQR of 1$^{\text{st}}$ principal strain could serve as a promising biomechanical signature for diagnosing pathological leaflets.

We organized the patient cohort into normal and diseased groups and compared the average areal and 1$^{\text{st}}$ principal strains (Fig.~\ref{fig:pediatric_MV_analysis}D). Neither group exhibited notable changes in leaflet area between the mid-diastolic and mid-systolic frames. In the normal cohort, the average areal strain was 0.034 $\pm$ 0.147 in the anterior leaflet and 0.029 $\pm$ 0.171 in the posterior leaflet. In the diseased cohort, the corresponding values were –0.038 $\pm$ 0.198 and 0.018 $\pm$ 0.354.  

The posterior leaflets exhibited higher 1$^{\text{st}}$ principal strain than the anterior leaflets in both normal and diseased cohorts. In the normal cohort, the average 1$^{\text{st}}$ principal strain was 0.262 $\pm$ 0.152 in the anterior leaflet and 0.337 $\pm$ 0.117 in the posterior leaflet. In the diseased cohort, the corresponding values were 0.223 $\pm$ 0.124 and 0.388 $\pm$ 0.356. No qualitative differences in 1$^{\text{st}}$ principal strain were observed between the two cohorts, as the inter-cohort differences for each leaflet were less than 0.05.

\subsubsection{Population study of strain patterns across the patient cohort}
The variations of MV morphology in this cohort and the associated strain patterns are shown in Fig.~\ref{fig:pediatric_MV_analysis}E. Within the spectrum, the MV with the smallest anterior–posterior annular diameter corresponded to -2 SD from the mean shape, whereas the largest corresponded to +2 SD. The MVs at both ends of the spectrum exhibited higher peak strains than the mean shape. At –2 SD, the MV showed uniformly elevated strain across the entire posterior leaflet.

Although differences in MV strain were modest, we observed a unidirectional increase in mean areal strain from –0.084 to –0.035 as MV morphology shifted from –2 to +2 SD of valve shapes. In contrast, both ends of the morphological spectrum exhibited higher average 1$^{\text{st}}$ principal strain than the mean shape: 0.162 at –2 SD and 0.191 at +2 SD, compared to 0.131 in the mean shape. Similar to the TV analysis, the SD of areal and 1$^{\text{st}}$ principal strain trended greater as valve geometries deviated further from the mean.

\subsection{Strain assessment of adult mitral valves (N=10)}~\label{sec:adult_MV_strain}

\begin{figure*}[h!]
\centering
\includegraphics[width=\textwidth]{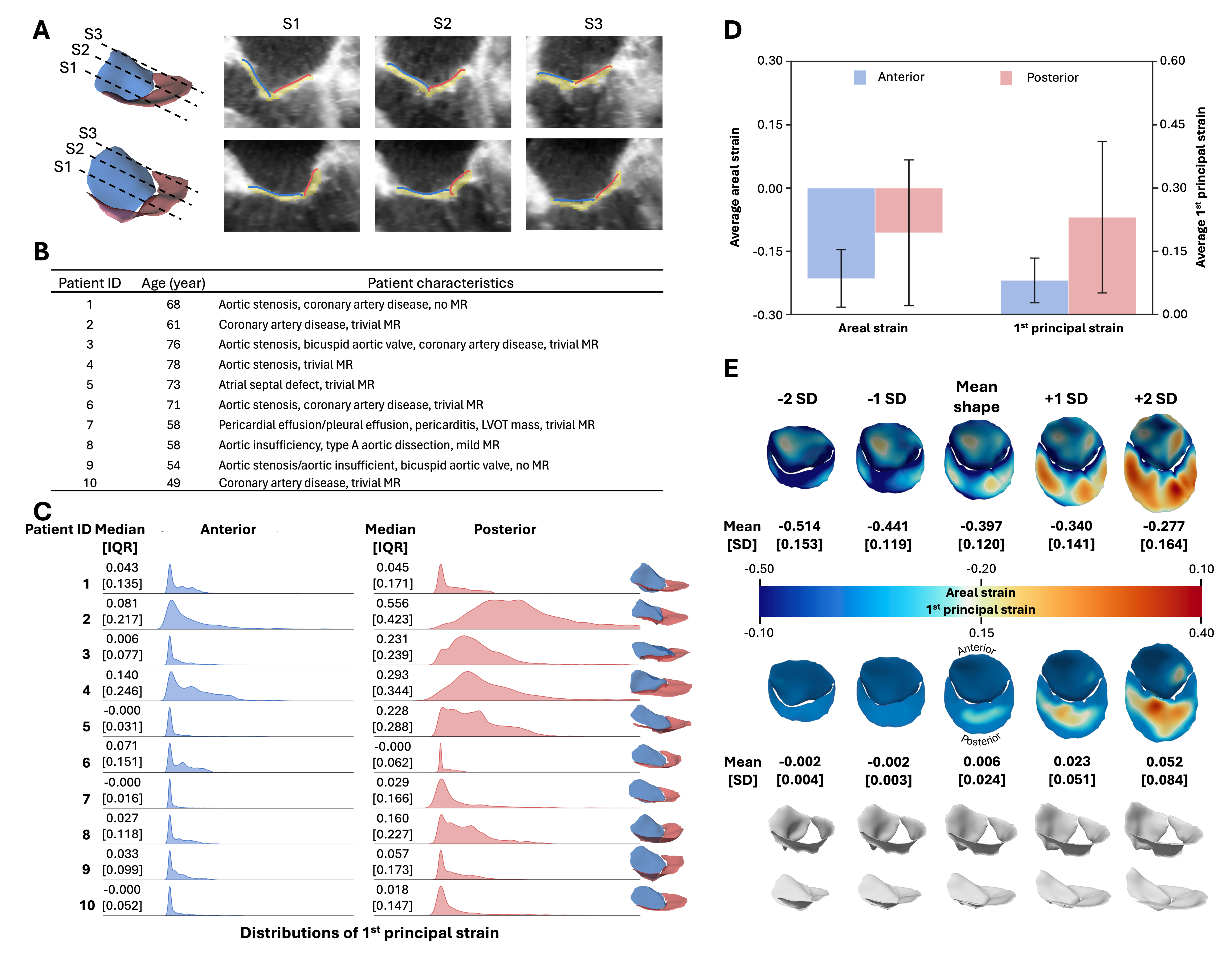}
\caption{\textbf{Adult mitral valve analysis (N=10).} (\textbf{A}) Visualization of the ground truth segmentation models and estimated atrial surfaces were superimposed onto 3D TEE images. (\textbf{B}) The study included adult patients aged 49 to 78 years at the time of image acquisition. All patients had normal MV function but presented with other cardiac conditions. (\textbf{C}) Narrow 1$^\text{st}$ principal strain distributions were observed in all MV leaflets. Further, the median and IQR of the strain distributions were below 0.5 across all 10 patients. (\textbf{D}) Both anterior and posterior leaflets exhibited negative areal strain, which may suggest substantial annular contraction between the mid-diastolic and mid-systolic frames in this cohort. For 1$^{\text{st}}$ principal strain, consistent with the pediatric MV study, the posterior leaflet showed higher strain than the anterior leaflet. (\textbf{E}) In the population-based strain analysis, average areal strain and 1$^{\text{st}}$ principal strain both increased with increasing anterior–posterior diameter.} \label{fig:adult_MV_analysis}
\end{figure*}

\subsubsection{Patient population}
This study cohort comprised 10 adult cardiac surgery patients, aged 49 to 78 years at the time of intra-operative image acquisition. All patients demonstrated normal MV function with no to mild mitral regurgitation (MR), though each had non-mitral cardiac pathology. Excellent feature tracking accuracy was demonstrated in all patients.  The ground truth segmentation and the estimated atrial leaflet surface overlaid onto 3D TEE images from two representative patients are presented in Fig.\ref{fig:adult_MV_analysis}A. Further, Fig.\ref{fig:adult_MV_analysis}B summarizes the ages and clinical diagnoses of the entire cohort. Unlike the previous two cohorts, none of the patients in this group were identified with billowing leaflets. 

\subsubsection{Strain characteristics on individual leaflet} 
Fig.~\ref{fig:adult_MV_analysis}C presents ridgeline plots comparing the 1$^{\text{st}}$ principal strain in the anterior and posterior leaflets. In most patients, the MV leaflets exhibited narrow strain distributions with low median and IQR values. An exception was Patient 2, whose posterior leaflet demonstrated a broader distribution with a median of 0.556 and an IQR of 0.423.

Fig.~\ref{fig:adult_MV_analysis}D compares the average areal and 1$^{\text{st}}$ principal strain of each leaflet across the patient cohort. Interestingly, negative average areal strain was observed in both leaflets (anterior: –0.215 $\pm$ 0.068; posterior: –0.107 $\pm$ 0.173). This may indicate a significant contraction of the annulus within the mitral valve in the adult population. However, further analysis is warranted to identify the causes of the negative average areal strain. For 1$^{\text{st}}$ principal strain, the posterior leaflet exhibited higher values than the anterior leaflet (0.231 $\pm$ 0.180 vs. 0.081 $\pm$ 0.053).

\subsubsection{Population study of strain patterns across the patient cohort}
Fig.~\ref{fig:adult_MV_analysis}E illustrates the spectrum of MV morphologies in this cohort along with their associated strain distributions. Consistent with the pediatric cohort, the valve with the shortest anterior–posterior annular diameter corresponded to –2 SD from the mean shape, whereas the longest corresponded to +2 SD. In this cohort, both average areal strain and 1$^\text{st}$ principal strain increased as the anterior–posterior annular diameter increased from –2 SD to +2 SD of valve shapes. In terms of strain heterogeneity, the SD of areal strain increased as the valve geometries deviated further from the mean shape. Conversely, the SD of the 1$^\text{st}$ principal strain increased undirectionally from -2 SD to +2 SD of the valve shape.

\section{Discussion}

\subsection{General comments}
In this study, we introduce an open-source, feature tracking-based framework for quantifying leaflet strain in atrioventricular valves. Implementation accuracy was verified using a FEA benchmark problem, where agreement was observed in both shape similarity and strain estimation, with Pearson's $r \geq 0.710$ ($p < 0.001$) and Spearman's $\rho \geq 0.746$ ($p < 0.001$) in both areal and 1$^\text{st}$ principal strain. Relative to BCPD~\cite{hirose_bayesian_2021} and classic CPD~\cite{myronenko_point_2010}, our method consistently achieved excellent performance across all evaluated metrics. To assess robustness to morphological variation, we applied the framework to three patient cohorts: (1) TVs in patients with HLHS, (2) pediatric MVs, and (3) adult MVs. The dataset spans normal to regurgitant valves and ages from neonates to the late seventies. This constitutes the first human analysis of tricuspid valve strain quantification comparing healthy and diseased cohorts, and the most heterogeneous patient population studied to date. Despite substantial variation in valve geometry across patients and cohorts, the same feature tracking parameters produced excellent shape agreement in all cases. This highlights the generalizability of the framework. Across cohorts, the median and IQR of the 1$^\text{st}$ principal strain distributions emerged as a potential marker of altered leaflet biomechanics, with elevated median and IQR values in valves with leaflet billow.

To provide a summary of our strain analysis for MVs, qualitative differences in the characteristics of 1$^\text{st}$ principal strain distributions between pediatric and adult MVs were observed. In pediatric valves, the distribution was broader with a higher IQR, whereas in adults, it was narrower and more peaked. In cohort-level PCA, both groups showed increasing strain IQR as valve geometries deviated from the mean shape. In adults, distinct patterns emerged in which strain in the posterior leaflet progressively increased and became more diffuse with increasing anterior–posterior diameter. In contrast, this pattern was not observed in the pediatric MV cohort. In our pediatric cohort (mean age 8.9 $\pm$ 4.9 years), the average $1^\text{st}$ principal strain was 0.262 in the anterior leaflet vs. 0.337 in the posterior leaflet in valves with normal morphology. Conversely, the average 1$^\text{st}$ principal strain reported in our adult cohort (mean age 64.6 $\pm$ 9.4 years) was markedly lower, with strains exhibiting 0.081 and 0.231, respectively. These preliminary findings are consistent with the general intuition that leaflet strain reduces with age, attributed to an increase in tissue stiffness and thickness.

\subsection{Comparison with existing literature}
Strain in TVs varies substantially across leaflets~\cite{mathur_tricuspid_2019, larue_tricuspid_nodate}. In our study, patients with mild or greater regurgitation exhibited greater average areal strain and 1$^{\text{st}}$ principal strain in the anterior leaflet, but lower values in the posterior leaflet, compared with the trivial to moderate cohort. Across both normal and diseased groups, the posterior leaflet consistently experienced the greatest stretch, followed by the anterior and septal leaflets. This pattern aligns with findings reported by Spinner et al.~\cite{spinner_effects_2012}. However, Mathur et al. identified the anterior leaflet as having the highest areal strain~\cite{mathur_tricuspid_2019}. Analysis at the individual leaflet level further showed that billowing leaflets tend to exhibit higher IQR in the 1$^{\text{st}}$ principal strain, suggesting greater strain heterogeneity. Strain analysis in larger cohorts is needed to verify these observations. Cohort-level PCA demonstrated that the normal valve, represented by the mean shape, has the lowest average strain and IQR. As valves deviate from this mean toward tethering or billowing phenotypes, both the average strain and the IQR increase.

In our MV studies, we observed higher average strain in the posterior leaflet compared with the anterior leaflet, consistent with the literature~\cite{el-tallawi_valve_2021}. This trend was evident in our 1$^\text{st}$ principal strain calculations across pediatric and adult populations, and in both normal and diseased valves. In adults, El-Tallawi et al.~\cite{el-tallawi_valve_2021} reported isotropic strain measurements in normal adult mitral valves and observed modest differences between anterior and posterior leaflets. In our adult cohort, posterior leaflet strain was 2.8 times higher than anterior leaflet strain, a more pronounced difference between leaflets than previously reported. We attribute this difference to variation in the cardiac phases selected for strain calculation. Prior work quantified strain over a narrower systolic interval, whereas our analysis spanned a broader interval from mid-diastole to mid-systole, which captures a larger deformation range and results in higher measured strain.

\subsection{Limitations and area for future work}
While we validate and demonstrate the ability of our method to accurately capture leaflet strains across diverse valve geometries, we were not powered to explore relationships between leaflet strain, geometry, and pathology;  our findings in this domain are exploratory. Several additional technical considerations may influence the interpretation of leaflet strain, particularly the occurrence of negative areal strain (i.e., stretch less than 1)~\cite{rego2022vivo, rego2022patient, nakhaei2026biomechanically}. From a mechanical perspective, leaflet area would generally be expected to increase during systole as pressurization induces tensile loading and tissue stretch, although localized contraction or complex folding of the leaflet may occur in specific regions. When computing leaflet area of the segmental atrial surface between diastole and systole, however, we observed an overall negative percent change in area. This behavior could potentially reflect imaging and reconstruction limitations in addition to any true physiological effects. In systolic 3DE, coaptation zones and locally folded regions are not sufficiently resolved (Fig.~\ref{fig:pediatric_TV_analysis}-~\ref{fig:adult_MV_analysis} panel A), which can preclude full capture of the native leaflet surface and lead to systematic underestimation of the true leaflet area. Because areal strain is derived directly from these surface measurements, such underestimation contributes to negative strain values. However, these effects arise from consistent constraints in image quality and segmentation and are therefore systematic across leaflets and patient populations, preserving the reproducibility of relative comparisons and regional trends. Although our pipeline is compatible with segmentations generated from multiple software platforms, differences in smoothing or surface regularization may introduce modest variability in local geometry and consequently in strain measurements. Continued improvements in image resolution and segmentation fidelity will be important for reducing these sources of bias and further enhancing quantitative accuracy of strain estimation.

Notably, tissue thickness also influences the rate of deformation (\textit{i.e.,} the extent to which the tissue stretches or strain rate). While our approach provided robust estimates of systolic valve deformation and strain, it did not account for spatial variation in leaflet thickness, which may influence local strain patterns. Further, factors such as signal dropouts in the echocardiographic images can introduce errors into the segmentation models. This error may subsequently degrading the accuracy of the final strain analysis. While are methods are fundamentally applicable to any 3D + time imaging modality, we only demonstrated application to echocardiography given the ready availability of images with sufficient temporal resolution. In our current analysis, we observed a promising trend suggesting that leaflet billowing is associated with an elevated median and IQR of 1$^\text{st}$ principal strain, a finding that appears consistent across both MVs and TVs. Validation in larger cohorts will be necessary to confirm this observation. Although our method allows the study of leaflet strain progression over time and the evaluation of postoperative leaflet strain, it cannot be used to predict surgical outcomes in advance without a target systolic surface. Research efforts aimed at improving the fidelity of FEA are critical to accurately emulate ground-truth physiological valvular dynamics and enable \textit{de novo} forward simulations to identify the optimal repair in an individual patient.

\subsection{Concluding remarks}
We have developed a novel point-based feature tracking approach for tracking leaflet motions across the diastole and systole frames. Our method provides a comparable strain estimation to that in FEA. Our approach was successfully applied to three groups of atrioventricular valves across pediatric and adult patients (a total of 31 cases). The cohorts encompassed a broad spectrum of valve anatomy, from healthy mitral valves to those with restricted leaflets and pronounced billowing. Although we did not identify a unifying strain pattern that clearly distinguishes healthy from diseased valves, we observed a promising trend in which a higher median and IQR of strain distribution appears to be a strong marker of leaflet billowing. With further validation, this metric could serve as a novel quantitative tool to enhance prognostication of valvular heart disease.

\section{Acknowledgments}

This work was supported by the Cora Topolewski Pediatric Valve Center at the Children's Hospital of Philadelphia, an Additional Ventures Single Ventricle Research Fund award, and the National Institutes of Health NHLBI K25 HL168235 (WW), NHLBI R01 HL153166 (MAJ), NHLBI R01 HL163202 (AMP), NIGMS R01 GM083925 (JAW).

\section*{CRediT authorship contribution statement}
\authcredits{Wensi Wu}{Conceptualization, Formal analysis, Funding acquisition, Investigation, Methodology,  Visualization, Writing - original draft}
\authcredits{Matthew Daemer}{Data curation, Writing – review \& editing}
\authcredits{Jeffrey A. Weiss}{Conceptualization, Funding acquisition, Writing – review \& editing}
\authcredits{Alison M. Pouch}{Conceptualization, Data curation, Funding acquisition, Writing – review \& editing}
\authcredits{Matthew A. Jolley}{Conceptualization, Funding acquisition, Writing – review \& editing}


\printbibliography

\end{document}